\documentstyle[pre,aps,preprint,eqsecnum]{revtex}
\begin{document}

\title{STABILITY LIMIT FOR MODE-I FRACTURE}

\author{Emily S. C. Ching* and J.S. Langer}

\address{Institute for Theoretical Physics\\
University of California\\
Santa Barbara, California 93106-4030}

\date{\today}

\maketitle

\begin{abstract} In order to study the stability of mode-I fracture, we
consider a crack moving along the centerline of a very wide strip and compute
its steady-state response to a small, spatially periodic shear stress. We find
that, in the presence of this perturbation, the crack remains very nearly
straight up to a critical speed $v_c$ of about $0.8\,v_R$, where $v_R$ is the
Rayleigh speed. Deviations from straight-line propagation are suppressed by a
factor proportional to $W^{-1/2}$, where $W$ is the width of the strip. At
$v_c$, however,  this suppression disappears and the steady-state crack follows
the wavy curve  along which the shear stress vanishes in the unbroken strip.
We interpret this  behavior as a loss of stability and discuss its implications
for a more complete  dynamical theory of fracture propagation.

\end{abstract}

\bigskip

*Present address: Department of Physics, The Chinese University of Hong Kong,\\
Hong Kong.

\newpage

\section{Introduction and Summary of Results}

The need for an analytic understanding of the dynamic stability of propagating
fracture has acquired special urgency in the last several years. This renewed
interest in an old subject is largely the result of some remarkable experiments
by Fineberg {\it et al}\cite{Texas} and also some very recent numerical
experiments by Abraham\cite{Abraham} in which similar instabilities are
observed.  Our purpose in this report is to outline a new approach  to the
theory of fracture stability and to present one small but, we believe,
important  result --- specifically, an upper limit to the speed of mode-I
cracks.

Both the laboratory and numerical experiments appear to indicate that
propagating mode-I fractures in a variety of materials encounter some kind of
oscillatory instability at velocities appreciably below the Rayleigh speed.
Motions at speeds higher than the onset velocity for the instability are
necessarily very dissipative; the oscillation radiates elastic energy and, as
seen in the experiments, induces extra deformation near the fracture surface.
It is not clear whether the onset velocity is an absolute upper bound for the
propagation speed or is simply a speed above which further acceleration becomes
much more difficult.  The experiments also seem to indicate that the
oscillation involves the direction of propagation and not just the speed.  That
is, the crack moves along a wiggly line through the solid, and the wiggles
usually are accompanied by  some form of sidebranching.

A basic idea of why such an instability might occur was suggested some years
ago by Yoffe\cite{Yoffe}.  (See Freund's book\cite{Freund} for a more modern
treatment.) The idea is conceptually quite simple.  Consider an idealized crack
in the form of a simple straight slit in a two-dimensional linear elastic
material and suppose --- without asking for a dynamic explanation --- that the
tip of this crack is moving forward at a constant speed $v$.  The singular
stress field near the tip can be obtained by transforming the equations of
elasticity into the moving frame.  The result, in effect, is a ``Lorentz
transformation'' of the field that deforms the pattern of longitudinal and
shear stresses.  Yoffe observed that, at some critical $v$ less than the
Rayleigh speed $v_R$, the maximum in the component of the stress that pulls the
fracture surfaces apart from each other shifts away from the forward direction.
In effect, the Lorentz-Fitzgerald contraction enhances the stress parallel to
the direction of crack motion. At speeds higher than the critical $v$, this
maximum occurs at an angle of about $60^{\circ}$ on either side of the forward
direction; it is tempting to speculate that the crack might undergo a
directional instability under these conditions.

The shortcoming of the Yoffe analysis is that it contains no consideration of
the forces required to keep the crack moving or any prediction of how the crack
actually responds to these forces.  A real crack cannot support the singular
stresses at the tip that are produced by a simple slit.  There must be some
mechanism by which these stresses are regularized, for example, by blunting of
the tip or by deformation of the material in its neighborhood.  One of the
present authors (E.S.C.C.) recently has studied these stresses in detail for
the model used here and has found that the maximum in the regularized stress at
the tip moves off axis at a crack velocity less than the Rayleigh
speed\cite{Ching}. In any case, Yoffe's singular stress cannot be, and indeed
is not, the actual stress acting at the physical tip of the crack.  Therefore
it is not clear how or whether the Yoffe stress can determine the dynamic
response of the system. It is that area of uncertainty that we address in this
paper.

The analysis presented here is limited in its purposes but is already quite
complicated.  It is useful, therefore, to start with a summary of our strategy
and main results.

We consider a crack moving along the centerline of an infinite elastic strip
occupying the region $(-\infty<x<+\infty, -W<y<+W)$ in the $x,y$ plane.  Far
ahead of the crack, the strip is uniformly strained by an amount
$\varepsilon_{xx}= \varepsilon_{\infty}^{(1)} $, $\varepsilon_{yy}=
\varepsilon_{\infty}^{(2)}$ and, for the moment, $\varepsilon_{xy}=0$.  From
the  beginning of the analysis, we assume that the half-width $W$ is very much
larger than any other length scale in the problem, thus we carry out most of
our calculations in the limit $W\to\infty$ and impose outgoing-wave boundary
conditions at large $|y|$.  However, there are several places where we need to
reintroduce the length $W$.  For example, the fully relaxed width of the crack
must scale like $W$, and the stress-intensity factor that characterizes the
forces transmitted to the crack tip is proportional to $\sqrt W$.  Our
technique for determining these W-dependences is based on an earlier paper in
this series\cite{BDL}, hereafter referred to as BDL.

The minimal ingredients of a phenomenological theory of fracture dynamics are
mechanisms for regularizing the stress singularity, for introducing a fracture
energy, and for dissipating energy.  Our results appear to be insensitive to
our choices for these ingredients, but we prefer to work within the context of
a specific dynamic model.  The particular model to be used here is a mode-I
generalization of the mode-III model described in Ref.(\cite{LN}), hereafter
referred to as LN, which contains a conventional Barenblatt cohesive zone and
an unconventional viscous dissipation on the fracture surface.  The most
remarkable feature of this model is that, when the dissipative force is
sufficiently strong, the crack creeps only very slowly at external stresses
just above the Griffith threshold and then, as a function of increasing applied
stress,  makes a rapid but smooth transition to propagation at the Rayleigh
speed.  This  transition occurs at a dissipation-dependent effective threshold.
The important  point for present purposes is that the structure and the
steady-state motion of the  crack are completely determined in this model;
there are no unphysical singularities  and there is no ambiguity in the
relationship between the applied force and the  response of the system.

So far as we know, no systematic investigation of the dynamic stability of
fracture has been reported before now.  The propagating fracture modes obtained
in the LN model, for example, are strictly steady-state solutions of the
equations of motion which may or may not have sufficient stability to be seen
in real or numerical experiments. In a conventional stability analysis, we
would start by linearizing the equations of motion about the steady-state
solutions and then would look to see whether the eigenstates of the resulting
linear operator grow or decay exponentially as functions of time.  To assure
ourselves of stability, we would have to determine that no member of the
complete set of eigenfunctions is a growing mode.  This would be a very
difficult program to carry out in the case of dynamic fracture because the
mathematics is very complicated.  However, even if this program could be
carried out rigorously and completely, it very likely would give us a wrong
answer. The reasons for failure of the eigenvalue method have been described
clearly in a recent paper by Trefethen {\it et al} \cite{Trefethen}.

The relevant complication is that the linear stability operator in this kind of
situation is non-normal, that is, it is not self-adjoint.  The lack of
normality occurs here because we are dealing with an externally driven system
and are linearizing about a steady-state solution in a moving frame of
reference. As a result of the non-normality of the linear operator,
realistically small perturbations may excite formally stable deformations
(linear combinations of non-orthogonal eigenmodes) that grow appreciably before
decaying at later times.  These transient growth modes may drive the system
into a nonlinear regime and thus produce behavior that is qualitatively
different from what one might expect on the basis of only the eigenvalue
analysis.  Effects of this kind have been studied so far primarily in fluids or
diffusive systems.  One example that is specially familiar to the present
authors is the Saffman-Taylor viscous finger in a Hele-Shaw cell. Here, the
steady-state finger is linearly stable according to conventional analysis, but
small perturbations at the tip may be greatly amplified before dying out as
they are advected down the finger.  A similar phenomenon occurs in dendritic
crystal growth where the amplified perturbations produce sidebranches.  Both of
the latter phenomena\cite{JSLSci} bear an intriguing but, we suspect, only
superficial resemblance to the fracture problem.

        With these difficulties in mind, we have chosen not to attempt a
conventional stability analysis.  Instead, we have adopted a strategy
that is both feasible for us and gives us
relevant, albeit limited, information. Specifically,
 we compute the steady-state
response of our system to a small ({\it i.e.} first order) external force that
produces a spatially oscillating shear stress along the $x$-axis:
\begin{equation}
\sigma_{xy}^{(ext)}(x,0)= 2\mu{\hat\varepsilon_m}\,e^{imx}, 
\end{equation}
where $\mu$ is the elastic modulus, $m$ is the wavenumber of the perturbation,
and $\hat\varepsilon_m$ is its amplitude.
(Because $\hat\varepsilon_m$ appears only in first order
throughout this analysis, we may conveniently use the complex representation
(1.1) with the understanding that physical quantities always are obtained by
taking real parts.)
In Section II we define $\sigma_{xy}$ in the entire $x,y$ plane
so that, in principle, it can be
the result of tractions applied at the edges of the
strip or of material irregularities near the centerline.
The goal of the calculation
is to compute the perturbed centerline $y=Y_{cen}(x)$ of the
resulting fracture to
first order in ${\hat\varepsilon_m}$, that is:
\begin{equation}
Y_{cen}(x) = \hat Y_m  \,e^{imx} \equiv
\hat\chi_Y(m,v)\,{\hat\varepsilon_m}\,e^
   {imx}.
\end{equation}
Here, $\hat\chi_Y$ is a complex steady-state response coefficient that depends
on the wavenumber $m$ and the average crack propagation speed $v$ which, in
turn, is determined by the applied strains $ \varepsilon_{\infty}^{(1)} $ and
$\varepsilon_{\infty}^{(2)} $. If $\hat\chi_Y$ diverges at some value of
 $v$, or
simply becomes anomalously
large there, then we conclude that the system undergoes some kind of dynamic
instability at that speed.

The wavy crack described by (1.2) is similar to that considered by
Gao\cite{Gao}, whose work draws on that of Cotterell and Rice\cite{CR}. It is
also reminiscent of the interesting quasistatic fracture patterns observed by
Yuse and Sano\cite{YS} but is, in fact, quite different because here we are
considering very fast, freely propagating fracture.  The technique of looking
at the  first-order response to small perturbations also has been used very
recently by  Rice {\it et al}\cite{RBK,PR}, who have studied the in-plane
stability of a fully three  dimensional crack.  Our work, like that of Gao,
focuses entirely on out-of-plane  deformations.

        Our strategy for computing $\hat\chi_Y(m,v)$ is to use the steady-state
techniques developed in LN to calculate terms up to first order in
${\hat\varepsilon_m}$.  As in LN, the crucial ingredient is the condition
that all stresses be nonsingular at the crack tip.  We start this calculation
by transforming the equations of elasticity
into a frame of reference moving in the negative $x$-direction at a speed such
that the tip of the crack is always at $x'=0$, that is, $x=x'+x_{tip}(t)$,
where
\begin{equation}
\dot x_{tip}(t)= - v - \hat v_m e^{-imvt}. 
\end{equation}
This transformation into a nonuniformly moving frame is absolutely essential
because it allows us to deal nonperturbatively with the various mathematical
singularities that occur at the crack tip.
Note that the first-order part of the
velocity $\dot x_{tip}$ oscillates with a frequency $mv$ and has an as yet
unknown complex amplitude $\hat v_m$ which, like $\hat Y_m  $, is
proportional to
${\hat\varepsilon_m}$:
\begin{equation}
\hat v_m \equiv \hat\chi_v(m,v)\,{\hat\varepsilon_m}. 
\end{equation}

        The next step is to write down formal solutions of the equations of
elasticity separately in the two regions of the $x,y$ plane above and below
$Y_{cen}(x)$, and then to evaluate the unknown coefficients that occur in those
solutions by imposing boundary conditions on the centerline.  Ahead of the
crack tip, the centerline is purely fictitious and the boundary conditions
are simply
statements that the stresses and displacements must be continuous there.
Behind the tip, on the other hand, these boundary conditions are the usual
statements about tractions on the fracture surfaces.  The result is a set of
Wiener-Hopf equations that can be solved for the unknown stresses ahead of
the tip and the unknown crack-opening displacements behind it.

        At zero'th order in ${\hat\varepsilon_m}$, the Wiener-Hopf
equation is the mode-I analog of the LN equation.  It acquires
precisely the same mathematical form
once the $W$-dependence is reinserted and, therefore, its solution can be
obtained simply by reinterpreting the variables that occur in LN.
The condition
that the zero'th order stress be nonsingular at the crack tip then
determines the average velocity $v$ uniquely as a function of
$\varepsilon_{\infty}^{(1)}$ and and $\varepsilon_{\infty}^{(2)}$.

At first order in ${\hat\varepsilon_m}$, the problem separates for reasons of
symmetry into two decoupled equations.  One of these involves only the
tangential displacement along the fracture surface, the shear stress ahead of
the crack tip, and the amplitude $\hat Y_m  $.  The second contains only the
normal stresses  and displacements and the amplitude $\hat v_m$.  Both
equations contain terms proportional to the zero'th order displacement.
Because these equations are decoupled, each can be solved by Wiener-Hopf
techniques.  Then the unknown amplitudes $\hat Y_m  $ and $\hat v_m$ can be
determined uniquely by requiring that the first-order shear and normal stresses
be nonsingular at the crack  tip.  By far the most important of these results
is the first, {\it i.e.} the one that pertains to the amplitude $\hat Y_m  $,
which describes deformations perpendicular to the direction of propagation.  It
is in this quantity that we expect a  large linear response to the perturbing
shear stress if the system undergoes some  kind of Yoffe instability. The
amplitude $\hat v_m$, on the other hand, describes a component of the response
--- the perturbed velocity in the $x$-direction --- that  should not be
particularly sensitive at first order to small excursions in the $y$-
direction.

        For the physically interesting case where the wavelength $2\pi/m$ is
much longer than the cohesive zone at the tip and much shorter than $W$, the
response coefficient $\hat\chi_Y$ can be written in the form:
\begin{equation}
\hat\chi_Y^{-1}(m,v)= -im\,\left[\Delta\epsilon_{\infty}+
\varepsilon_{\infty}^{(2)}(-imW)^{1/2} \,{\cal D}_Y(v)
\right], 
\end{equation}
where $\Delta\epsilon_{\infty}= \varepsilon_{\infty}^{(2)}-
\varepsilon_{\infty}^{(1)}$.  The quantity ${\cal D}_Y(v)$ is a real,
$m$-independent function of $v$ that is positive at small $v$ and decreases
through zero at a value of $v=v_c$ (depending on the Poisson ratio) somewhat
smaller than $v_R$.

Equation (1.5) has a number of remarkable implications.  When ${\cal D}_Y=0$,
the resulting spatial oscillation with wavenumber $m$ and amplitude $\hat Y_m
=i{\hat\varepsilon_m}/m\,\Delta\epsilon_{\infty}$ is the curve along    which
the shear stress vanishes in the absence of a crack.  When a crack is  present,
however, and ${\cal D}_Y(v)$ is of order unity, the factor $\sqrt W$ implies
that the right-hand side of (1.5) is completely dominated by the second term in
the square brackets and that the crack is strongly confined to the $x$-axis.
This confinement ceases abruptly when ${\cal D}_Y$ vanishes at $v=v_c$, and
therefore we may interpret $v_c$ as a velocity at which the crack undergoes
some loss of stability.  Note, however, that $\hat\chi_Y$ cannot diverge at
$v_c$ or at any other velocity; the two terms on the right-hand side of (1.5)
have different phases and cannot cancel each other.  Thus the system remains
formally stable against the perturbation (1.1) in the conventional sense.

        We offer some further remarks and speculations about Eq.(1.5) in the
concluding Section VII of this paper.  The next several Sections are devoted to
the details of our mathematical development.

\section{Definitions and Basic Equations}

For completeness, and in order to establish our notation, we start by writing
the equations of motion for an isotropic elastic material.  We use the elastic
potentials $\Phi(x,y)$ and $\Psi(x,y)$.  In terms of these functions, the
displacements $u_x(x,y)$ and $u_y(x,y)$ are:
 \begin{equation}
u_x={\partial\Phi\over\partial x} + {\partial\Psi\over\partial y};\qquad
u_y={\partial\Phi\over\partial y} - {\partial\Psi\over\partial x}; 
\end{equation} and the stresses are: \begin{mathletters} \begin{eqnarray}
\Sigma_{xx}\equiv {\sigma_{xx}\over 2\mu} &&= \left(\kappa\over
2\right){\partial^2\Phi\over\partial x^2} +\left({\kappa\over 2}-
1\right){\partial^2\Phi\over\partial y^2} +{\partial^2\Psi\over\partial
x\partial y};\\ \Sigma_{yy}\equiv {\sigma_{yy}\over 2\mu} &&=
\left({\kappa\over 2}- 1\right){\partial^2\Phi\over\partial x^2}
+\left(\kappa\over 2\right){\partial^2\Phi\over\partial y^2}
-{\partial^2\Psi\over\partial x\partial y};\\ \Sigma_{xy} \equiv
{\sigma_{xy}\over 2\mu} &&= {\partial^2\Phi\over\partial x\partial y} + {1\over
2}{\partial^2\Psi\over\partial y^2} - {1\over 2}{\partial^2\Psi\over\partial
x^2}. \end{eqnarray} \end{mathletters}
Throughout all of this analysis, we use
the symbol $\Sigma$ to denote stresses measured in units $2\mu$. The parameter
$\kappa$ is the square of the ratio of the longitudinal to transverse sound
speeds, which is a function of the Poisson ratio $\nu$: \begin{equation}
{\kappa\over 2} = \cases {1-\nu\over 1-2\nu,&plane strain;\cr 1\over
1-\nu,&plane stress.\cr}

        The functions $\Phi$ and $\Psi$ can be written in the form:
\begin{mathletters}
\begin{eqnarray}
\Phi(x,y,t)&&= {1\over 2}\left(\varepsilon_{\infty}^{(1)} x^2 +
\varepsilon_{\infty}^{(2)} y^2\right)+ {{\hat\varepsilon_m}\,y\over
im}\, e^{imx} + \phi(x,y,t),\\
\Psi (x,y,t)&&= \psi(x,y,t),
\end{eqnarray}
\end{mathletters}where $\phi$ and $\psi$ satisfy the wave equations:
\begin{equation}
\ddot\phi=\kappa\nabla^2\phi;\qquad\ddot\psi=\nabla^2\psi.
\end{equation}
We have scaled the time $t$ so that the transverse wave speed is unity.

        The first two terms on the right-hand side of (2.4a) produce,
respectively, the uniform tensile stress that drives the crack and the
oscillating shear stress (1.1) that perturbs the rectilinear motion.
The latter already is written in an approximate form that is legal
because we consider only first-order deviations of the crack away from
the $x$-axis, and also because we do not need to be specific about the
mechanism that causes the perturbing shear stress.  If this stress is
produced by tractions at the edges of the strip, for example, a correct
form for this term would be the harmonic function
$({\hat\varepsilon_m}/im^2)\,\sinh (my)\,\exp(imx)$; but the exponential
growth at large $y$ is not necessary and, in any case, is irrelevant for
our purposes.  The fields $\phi$ and $\psi$ are the changes in $\Phi$
and $\Psi$ caused by the presence of the crack.  So long as we do not
immediately look for static solutions by setting $v=0$, we may use the
decoupled wave equations (2.5).  We must be careful, however, in taking the
limit $v\to 0$.

        The next step is to transform the wave equations (2.5) into a frame
of reference moving with the tip of the crack and to look for steady-state
solutions in this frame.  We write $\phi$ and $\psi$ in the form
$\phi (x - x_{tip},y,t), \ \psi(x-x_{tip},y,t)$ so that the tip is always
at $x'=x-x_{tip}=0$. Then, for
simplicity, we redefine $x'\to x$.  The transformed equations are:
\begin{mathletters}
\begin{eqnarray}
\ddot\phi-2\dot x_{tip}{\partial\dot\phi\over\partial x} + \dot x_{tip}^2
{\partial^2\phi\over\partial x^2} - \ddot x_{tip} {\partial\phi\over\partial x}
&&= \kappa{\partial^2\phi\over\partial x^2} +
\kappa{\partial^2\phi\over\partial y^2 }; \\
\ddot\psi-2\dot x_{tip}{\partial\dot\psi\over\partial x} + \dot x_{tip}^2
{\partial^2\psi\over\partial x^2} - \ddot x_{tip} {\partial\psi\over\partial x}
&&= {\partial^2\psi\over\partial x^2} + {\partial^2\psi\over\partial y^2}.
\end{eqnarray}
\end{mathletters}

        At this point it is convenient to separate the problem into parts
which are zero'th and first order in the perturbation ${\hat\varepsilon_m}$.
We do this by using (1.3) and by noting that the only explicit time
dependence remaining in $\phi$ and $\psi$ must be a first-order oscillation
of frequency $mv$.  Therefore we write:
\begin{mathletters}
\begin{eqnarray}
\phi(x,y,t)&&=\phi_0(x,y)+\phi_1(x,y)\,e^{-imvt}\\
\psi(x,y,t)&&=\psi_0(x,y)+\psi_1(x,y)\,e^{-imvt}
\end{eqnarray}
\end{mathletters}At zero'th order, $\phi_0$ and $\psi_0$ satisfy:
\begin{equation}
\beta_l^2{\partial^2\phi_0\over\partial x^2} +
{\partial^2\phi_0\over\partial y^2} =0;\qquad
\beta_t^2{\partial^2\psi_0\over\partial x^2} +
{\partial^2\psi_0\over\partial y^2} =0;
\end{equation}
where
\begin{equation}
\beta_l^2\equiv 1 - {v^2\over\kappa};\qquad \beta_t^2\equiv 1-v^2. 
\end{equation}
The first-order equations are:
\begin{mathletters}
\begin{eqnarray}
\beta_l^2{\partial^2\phi_1\over\partial x^2}+
{\partial^2\phi_1\over\partial y^2} +
{2imv^2\over\kappa}{\partial\phi_1\over\partial x} + {m^2v^2\over\kappa}\phi_1
&&= {2v \hat v_m \over\kappa}{\partial^2\phi_0\over\partial x^2} - {imv
\hat v_m \over\kappa}
{\partial\phi_0\over\partial x} \\
\beta_t^2{\partial^2\psi_1\over\partial x^2}+
{\partial^2\psi_1\over\partial y^2}
+ 2imv^2{\partial\psi_1\over\partial x} + m^2v^2\psi_1
&&= 2v\hat v_m {\partial^2\psi_0\over\partial x^2} - imv \hat v_m
{\partial\psi_0\over\partial x}
\end{eqnarray}
\end{mathletters}

        The general solutions of (2.8) are:
\begin{mathletters}
\begin{eqnarray}
\phi_0^{[\pm]}(x,y)&&=\int{dk\over 2\pi}\,\hat\phi_0^{[\pm]}(k)\,
e^{\mp\beta_l|k|y+ikx}, \\
\psi_0^{[\pm]}(x,y)&&=\int{dk\over 2\pi}\,\hat\psi_0^{[\pm]}(k)\,
e^{\mp\beta_t|k|y+ikx}
\end{eqnarray}
\end{mathletters}where the superscripts in square brackets $[\pm]$ mean
that the fields pertain to regions above or below the centerline,
$y>Y_{cen}(x)$ or $y<Y_{cen}(x)$.  Symbols with hats such as
${\hat\varepsilon_m}$ or $\hat\phi_0^{[\pm]}(k)$, as always
throughout this paper, denote Fourier amplitudes.  Note that we include
only decaying exponentials in (2.11), which means that we are immediately
going to the limit in which the width of the strip $W$ is very large.
We shall return to this point shortly.  Using (2.11) to evaluate the
inhomogeneous terms on the right-hand sides of (2.10), we find first-order
solutions of the form:
\begin{mathletters}
\begin{eqnarray}
\phi_1^{[\pm]}(x,y) &&= \int {dk\over 2\pi}
\left[\hat\phi_1^{[\pm]}(k)\, e^{\mp q_ly} +
{\hat v_m k\over mv}\hat\phi_0^{[\pm]}(k)\,
e^{\mp\beta_l|k|y} \right]\,e^{ikx}, \\
\psi_1^{[\pm]}(x,y) &&= \int {dk\over 2\pi} \left[\hat\psi_1^{[\pm]}(k)\,
e^{\mp q_ty} + {\hat v_m k\over mv}\hat\psi_0^{[\pm]}(k)\,e^{\mp\beta_t|k|y}
\right]\,e^{ikx},
\end{eqnarray}
\end{mathletters}where
\begin{equation}
q_l^2=k^2-{v^2\over\kappa}(k-m)^2,\qquad q_t^2=k^2-v^2(k-m)^2.
\end{equation}
Again, we include only decaying exponentials, so that we choose the real
parts of $q_l$ and $q_t$ to be non-negative.  For this oscillatory part
of the field, we also must specify outgoing-wave conditions; that is,
we must choose the imaginary parts of $q_l$ and $q_t$ to be non-positive.

        We now use the above expressions to evaluate the displacements and
stresses along the centerline $Y_{cen}(x)$.  We must separate these vector
and tensor quantities into components which are normal and tangential to
the local orientation of the crack.  To first order in $Y_{cen}$, the
normal displacement is:
\begin{eqnarray}
\nonumber
&& u_N^{[\pm]}\left(x,Y_{cen}\right)\cong u_y^{[\pm]}(x,0)
+ \left.{\partial u_y^{[\pm]}\over\partial y}\right|_{y=0}Y_{cen}(x)
- u_x^{[\pm]}(x,0) {dY_{cen}\over dx} \\
\nonumber
&& =\left[{\partial \Phi^{[\pm]}\over\partial y}-{\partial\Psi^{[\pm]}
\over\partial x}
\right]_{y=0} \\
&&+\left[\left({\partial^2 \Phi^{[\pm]}\over\partial y^2} -
{\partial^2\Psi^{[\pm]}\over\partial x\partial y}\right)
-im \left({\partial \Phi^{[\pm]}\over\partial x} +
{\partial\Psi^{[\pm]}\over\partial y}\right)\right]_{y=0}
\hat Y_m  \,e^{imx-imvt}.
\end{eqnarray}
We need similar expressions for the shear, that is, the
tangential displacement:
\begin{eqnarray}
\nonumber
&& u_S^{[\pm]}\left(x,Y_{cen}\right)\cong u_x^{[\pm]}(x,0)+ \left.
{\partial u_x^{[\pm]}\over\partial y}\right|_{y=0}Y_{cen}(x) +
u_y^{[\pm]}(x,0) {dY_{cen}\over dx} \\
\nonumber
&& =\left[{\partial \Phi^{[\pm]}\over\partial x}+
{\partial\Psi^{[\pm]}\over\partial y} \right]_{y=0} \\
&&+\left[\left({\partial^2 \Phi^{[\pm]}\over\partial x \partial y} +
{\partial^2\Psi^{[\pm]}\over\partial y^2}\right)
+ im \left({\partial \Phi^{[\pm]}\over\partial y} -
{\partial\Psi^{[\pm]}\over \partial x}\right)\right]_{y=0}\hat Y_m  \,
e^{imx-imvt};
\end{eqnarray}
for the normal stress:
\begin{eqnarray}
\nonumber
\Sigma_N^{[\pm]}&& \left(x,Y_{cen}\right)\cong
\Sigma_{yy}^{[\pm]}\left(x,Y_{cen}\right)-2\Sigma_{xy}^{[\pm]}(x,0)
{dY_{cen}\over dx} \\
\nonumber
&&\cong \left[\left({\kappa\over 2} -1\right)
{\partial^2 \Phi^{[\pm]}\over\partial x^2}
+\left({\kappa\over 2}\right){\partial^2 \Phi^{[\pm]}\over\partial y^2}-
{\partial^2\Psi^{[\pm]}\over\partial x\partial y}\right]_{y=0} \\
&&+\left[\left({\kappa\over 2} -1\right)
{\partial^3 \Phi^{[\pm]}\over\partial x^2 \partial y}+
\left({\kappa\over 2}\right) {\partial^3 \Phi^{[\pm]}\over\partial y^3}-
{\partial^3\Psi^{[\pm]}\over\partial x\partial y^2}\right]_{y=0}\hat Y_m \,
e^{imx-imvt};
\end{eqnarray}
and for the shear stress:
\begin{eqnarray}
\nonumber
\Sigma_S^{[\pm]}\left(x,Y_{cen}\right)
&& \cong \Sigma_{xy}^{[\pm]}\left(x,Y_{cen}\right)
+\left(\Sigma_{yy}^{[\pm]}(x,0)-\Sigma_{xx}^{[\pm]}(x,0)\right)
{dY_{cen}\over dx} \\
\nonumber
&&\cong\Biggl[{\partial^2 \Phi^{[\pm]}\over\partial x\partial y}+{1\over
2}{\partial^2\Psi^{[\pm]}\over\partial y^2}-{1\over
2}{\partial^2\Psi^{[\pm]}\over\partial x^2}\Biggr]_{y=0}
+\Biggl[\Bigl({\partial^3 \Phi^{[\pm]}\over\partial x \partial y^2}
+{1\over 2}{\partial^3 \Psi^{[\pm]}\over\partial y^3}
-{1\over 2}{\partial^3 \Psi^{[\pm]}\over\partial x^2\partial y}\Bigr)\\
&& +im\Bigl({\partial^2 \Phi^{[\pm]}\over\partial y^2}
-{\partial^2 \Phi^{[\pm]}\over\partial x^2}
-2{\partial^2\Psi^{[\pm]}\over\partial x\partial y}\Bigr)\Biggr]_{y=0}
\hat Y_m  \, e^{imx-imvt}.
\end{eqnarray}
In the second form of (2.16), we have used the fact that
$\Sigma_{xy}^{[\pm]}(x,0)$ is automatically of first order in
${\hat\varepsilon_m}$ and therefore
makes only a second-order contribution to
$\Sigma_N^{[\pm]}\left(x,Y_{cen}\right)$.

        To complete this introductory presentation, we need to specify
the mode-I LN model.  Define the relative normal and shear displacements:
\begin{equation}
U_{\alpha}(x)\equiv {1\over 2}\left[u_{\alpha}^{[+]}(x,Y_{cen})-
u_{\alpha}^{[-]}(x,Y_{cen})\right];
\end{equation}
where $\alpha = N,S$. We also need:
\begin{equation}
\delta\Sigma_{\alpha}(x) \equiv
{1\over2}\left[\Sigma_{\alpha}^{[+]}(x,Y_{cen}) -
\Sigma_{\alpha}^{[-]}(x,Y_{cen})\right], 
\end{equation}
and
\begin{equation}
\bar{\Sigma}_{\alpha}(x)\equiv {1\over 2}
\left[\Sigma_{\alpha}^{[+]}(x,Y_{cen})
+\Sigma_{\alpha}^{[-]}(x,Y_{cen})\right].
\end{equation}
The dynamic content of the model is contained in the statement that, within the
fractured region $x>0$,
\begin{equation}
\bar{\Sigma}_N(x) =\Sigma_c\{U_N(x)\}-{\eta\over 2\mu}{\partial^2\dot
U_N\over\partial x_{cen}^2}. 
\end{equation}
Here, $\Sigma_c\{U_N\}$ is the Barenblatt cohesive stress and $\eta$ is the
coefficient of viscous dissipation. The subscript on $x_{cen}$ means that the
second derivative is to be taken with respect to distance along the centerline.
In principle, the viscous forces should be applied separately to both fracture
surfaces, but this complication is irrelevant for our purposes.  We also must
specify that no cohesive or viscous forces produce shear stresses on the
fracture surfaces:
\begin{equation}
\bar{\Sigma}_S(x)=0,\qquad x>0; 
\end{equation}
By definition,
\begin{equation}
U_{\alpha}(x)=0,\qquad x<0. 
\end{equation}
Finally,
\begin{equation}
\delta\Sigma_{\alpha}(x)=0
\end{equation}
for all $x$ and both modes $\alpha$.  For $x<0$, this is just the condition
that
the stresses be continuous in the unbroken region. For $x>0$, (2.24) is a
statement about the symmetry of the forces introduced in (2.21).

\section{Zero'th Order Calculations}

        We look first at the solution of the steady-state problem for the
case in which the perturbation ${\hat\varepsilon_m}$ vanishes.  Many details
of this solution have been presented in LN.

        The zero'th order part of $U_N(x)$, which we denote by $U_{N0}(x)$,
is obtained from the first term on the right-hand side of (2.14) and the
zero'th order solutions (2.11).  In analogy to (2.19) and (2.20), we define:
\begin{mathletters}
\begin{equation}
\delta\hat\phi_n\equiv {1\over
2}\left[\hat\phi_n^{[+]}-\hat\phi_n^{[-]}\right];
    \qquad
\hat{\bar\phi}_n\equiv {1\over2}\left[\hat\phi_n^{[+]}+\hat\phi_n^{[-]}\right];
\end{equation}
and
\begin{equation}
\delta\hat\psi_n\equiv {1\over 2}
\left[\hat\psi_n^{[+]}-\hat\psi_n^{[-]}\right] ;\qquad
\hat{\bar \psi}_n\equiv {1\over 2}\left[\hat\psi_n^{[+]}
+\hat\psi_n^{[-]}\right];
\end{equation}
\end{mathletters}where the quantities $\hat\phi_n^{[\pm]}$ and
$\hat\psi_n^{[\pm]}$ are the Fourier amplitudes defined in (2.11)
and (2.12) and the subscript $n = 0,1$ denotes zero'th or first order
quantities. We find:
\begin{equation}
\hat{U}_{N0}(k)= -\beta_l|k|\hat{\bar{\phi}}_0(k)-ik\,\delta\hat{\psi}_0(k).
\end{equation}
Similarly,
\begin{equation}
\hat{\bar{\Sigma}}_{N0}(k) = 2\pi\Sigma_{\infty}\delta(k) +
k^2\beta_0^2\hat{\bar{\phi}}_0(k)+i\beta_tk|k|\,\delta\hat{\psi}_0(k);
\end{equation}
where
\begin{equation}
\Sigma_{\infty}\equiv \left(\kappa\over2\right)\varepsilon_{\infty}^{(2)}
+\left({\kappa\over2}- 1\right)\varepsilon_{\infty}^{(1)}, 
\end{equation}
and
\begin{equation}
\beta_0^2\equiv 1-{v^2\over 2}.
\end{equation}
We also need the symmetry and continuity conditions:
\begin{equation}
\delta{\hat{\Sigma}}_{S0}(k)=-i\beta_lk|k|\hat{\bar{\phi}}_0(k) +
k^2\beta_0^2\delta\hat{\psi}_0(k)=0;
\end{equation}
\begin{equation}
\hat{U}_{S0}(k)=ik\delta\hat{\phi}_0(k)-\beta_t|k|\hat{\bar{\psi}}_0(k)=0;
\end{equation}
and
\begin{equation}
\delta\hat{\Sigma}_{N0} = k^2\beta_0^2\delta\hat{\phi}_0
+i\beta_tk|k|\hat{\bar{\psi}}_0 =0.
\end{equation}

        Equations (3.7) and (3.8) immediately tell us that
\begin{equation}
\delta\hat{\phi}_0(k)=\hat{\bar{\psi}}_0(k)=0.
\end{equation}
Elimination of  $\hat{\bar{\phi}}_0(k)$ and $\delta\hat{\psi}_0(k)$ from
(3.2), (3.3) and (3.6) produces a relation between
$\hat{\bar{\Sigma}}_{N0}(k)$ and $\hat{U}_{N0}(k)$ which can be used to
evaluate the left-hand side of (2.21). The result is:
\begin{equation}
-\left(\hat{F}(k)+ i\,{\eta\over 2\mu}
\,v\,k^3\right)\,\hat{U}_{N0}^{(+)}(k)=\hat\Sigma_c^{(+)}(k) +
 \hat\Sigma_{N0}^{(-)}(k)-2\pi\Sigma_{\infty}\delta(k).
\end{equation}
Here,
\begin{equation}
\hat{F}(k)=b(v)|k|,\qquad  b(v)={2\over\beta_lv^2}(\beta_l\beta_t-\beta_0^4);
\end{equation}
and $\hat\Sigma_{N0}^{(-)}$ is the Fourier transform of the unknown zero'th
order stress in the unbroken region $x<0$.  As in previous papers, we have
introduced the superscripts in parentheses $(\pm)$ to indicate functions that
have singularities only in the upper (or lower) half $k$-planes because they
are Fourier transforms of functions that are nonzero only on the positive (or
negative) $x$-axis. In (3.11), note that $b(v)$ vanishes at the Rayleigh speed:
$b(v_R)=0$.

        It is at this point that we must pay attention to our assumption
that the width of the strip $W$ is much larger than any other length in
the problem.  In the Fourier representation, that condition is $kW\gg1$.
The one length scale that does not quite satisfy this condition is the
width of the fully open crack, which scales like $W$ and which plays an
important role in our analysis. In particular, we need an approximation
for the function $\hat F(k)$ for small $k$. We can compute the exact
value of $\hat F(0)$ by looking at the case where the crack is
fully open everywhere so that $U_{N0}=\varepsilon_{\infty}^{(2)} W$ and
$\bar\Sigma=\Sigma_{\infty}-\hat F(0)U_{N0}=0$. Thus,
\begin{equation}
\hat F(0)={\Sigma_{\infty}\over\varepsilon_{\infty}^{(2)} W}
\equiv {\kappa'\over 2W},
\end{equation}
where $\kappa'\equiv \kappa+(\kappa-2)(\varepsilon_{\infty}^{(1)}/
\varepsilon_{\infty}^{(2)})$.  Then, in analogy
to BDL, we make the simple interpolation:
\begin{equation}
\hat F(k)\approx
\left[\left(\kappa'\over2W\right)^2+b^2(v)k^2\right]^{1\over2}.
\end{equation}

        With (3.13), Eq.(3.10) becomes identical in form to Eq.(3.3) in LN
apart from a difference in the way we have subtracted the constant stress
$\Sigma_{\infty}$.  Thus we can obtain the complete zero'th-order solution
of our present model simply by reinterpreting the variables in LN. For
completeness, we list a few of these transformations.  Variables with tildes
are those which appear (without tildes) in LN.  For example, the applied
stress is $\tilde\varepsilon_{\infty}=\Sigma_{\infty}$. The Wiener-Hopf
kernel in LN has the form:
\begin{equation}
\hat{\tilde Q}(\tilde k)=\sqrt{1+\tilde k^2}+i\tilde\nu k^3.
\end{equation}
where the Fourier variable is $\tilde k=[2Wb(v)/\kappa']\,k$ and
\begin{equation}
\tilde\nu = \left(\kappa'\over 2W\right)^2{\eta v\over 2\mu
b^3(v)}.
\end{equation}
We already see that the function $b(v)$ replaces $\tilde\beta(v)=\sqrt{1-v^2}$,
thus the Rayleigh speed plays much the same role that the sound speed played
in LN.  The important transition from creep to fast fracture is described by an
equation of the form:
\begin{equation}
{v\over b^3(v)}=\left(K_{\infty}\over K_{eff}\right)^{12}, 
\end{equation}
where $K_{\infty}\sim \Sigma_{\infty}\sqrt{W}$ is the stress intensity factor
and $K_{eff}$ is the same  $W$-independent function of the viscosity $\eta$
and the cohesive stress that appears in LN.

        We need one explicit zero'th order result for later use, specifically,
an expression for $U_{N0}(x)$ for values of $x$ well outside the cohesive zone.
This is:
\begin{equation}
U_{N0}(x)\approx
W\varepsilon_{\infty}^{(2)}\int_0^\xi{d\xi'\over\sqrt{\pi\xi'}}
   e^{-\xi'};  \qquad
\xi={\kappa'x\over2Wb(v)}. 
\end{equation}
This result is almost obvious; it is a simple interpolation between the
$\sqrt x$ behavior near the tip and the constant displacement
$U_{N0}\approx W\varepsilon_{\infty}^{(2)}$ as $x\to\infty$, and it
embodies our observation following (3.14) that the natural scale of distance
is $2Wb(v)/\kappa'$.  This result can be
derived more systematically from (4.17) in LN by using (4.1), (4.14), and the
large-$\xi$ versions of (5.8) and (5.11) in the latter paper.

\section{First-Order Calculations: Wiener-Hopf Solutions for the Antisymmetric
Components}

        The first-order calculation separates naturally into parts for which
the
displacements are symmetric and antisymmetric on reflection about the $x$ axis.
It is the antisymmetric part that determines the centerline $Y_{cen}(x)$ and
which, therefore, is of most interest here.

        From (2.15), (2.11) and (2.12), we compute the first-order shear
displacement:
\begin{eqnarray}
\nonumber
\hat U_{S1}^{(+)}(k)&&=ik\,\delta\hat\phi_1(k)-q_t\,\hat{\bar\psi}_1(k) \\
&&-\biggl\{i\beta_lk|k-m|\,\hat{\bar\phi}_0(k-m)
-(k-m)\Bigl[k-v^2(k-m)\Bigr] \delta\hat\psi_0(k-m)\biggr\}\hat Y_m  .
\end{eqnarray}
Similarly, from (2.16) and (2.24):
\begin{eqnarray}
\nonumber
\delta\hat\Sigma_{N1}(k)&&=q_0^2\,\delta\hat\phi_1(k)
+ikq_t\,\hat{\bar\psi}_1(k) \\
&&-\biggl\{\beta_0^2\beta_l|k-m|^3\,\hat{\bar\phi}_0(k-m)
+i\beta_t^2(k-m)^3\,\delta\hat\psi_0(k-m)\biggr\}\hat Y_m  =0;
\end{eqnarray}
where
\begin{equation}
q_0^2\equiv k^2-{v^2\over 2}(k-m)^2.
\end{equation}
Finally, from (2.17):
\begin{eqnarray}
\nonumber
\hat{\bar\Sigma}_{S1}^{(-)}(k)
&&=-ikq_l\,\delta\hat\phi_1(k)+q_0^2\,\hat{\bar\psi}_1(k)
+\biggl\{i(k-m)^2\Bigl(k+m-{v^2k\over\kappa}\Bigr)\hat{\bar\phi}_0(k-m) \\
&&-\beta_t(k-m)|k-m|\Bigl[k+m-{v^2\over2}(k-m)\Bigr]\delta\hat\psi_0(k-m)
\biggr\}\hat Y_m   +2\pi\,\hat E_m\delta(k-m),
\end{eqnarray}
where
\begin{equation}
\hat E_m\equiv {\hat\varepsilon_m}+im \Delta\epsilon_{\infty}\hat Y_m .
\end{equation}
Note that $\hat{\bar\phi}_1$, $\delta\hat\psi_1$, and $\hat v_m$
do not appear in
these equations.  The superscripts $(\pm)$ remind us that the relative shear
displacement vanishes ahead of the crack and the shear stress vanishes on the
fracture surfaces, thus $\hat U_{S1}^{(+)}(k)$ and
$\hat{\bar\Sigma}_{S1}^{(-)}(k)$ have singularities, respectively,
only in the upper and lower half $k$-planes.

        Eliminating $\delta\hat\phi_1$ and $\hat{\bar\psi}_1$ from the above
equations, and using (3.2) and (3.6) to evaluate $\hat{\bar\phi}_0$ and
$\delta\hat\psi_0$ in terms of $\hat U_{N0}^{(+)}$, we find an equation of the
form:
\begin{equation}
\hat{\bar\Sigma}_{S1}^{(-)}(k)=2\pi\,\hat E_m\delta(k-m)
-\hat G_S(k)\,\hat U_{S1}^{(+)}(k)+\hat L_S(k)\,
\hat U_{N0}^{(+)}(k-m)\,\hat Y_m  ,
\end{equation}
where the Wiener-Hopf kernel --- the analog of $\hat F(k)$
in (3.10) --- is:
\begin{equation}
\hat G_S(k)={2\over q_tv^2(k-m)^2}\Bigl(k^2q_lq_t-q_0^4\Bigr)
\end{equation}
and
\begin{eqnarray}
\nonumber
\hat L_S(k)=&&
{2\,ikq_l\over v^2}\Bigl[\Bigl({m\over k-m}\Bigr)^2-\beta_0^2\Bigr]
+{2\,ikq_0^2\over q_tv^2}\Bigl[1-\Bigl({m\over k-m}\Bigr)^2
-v^2\Bigl({k-m\over k}\Bigr)\Bigr] \\
&&+\Bigl({2\,i\over v^2}\Bigr)\,|k-m| \Bigl[{\beta_0^2\over \beta_l}
\Bigl(\beta_l^2k+m\Bigr)
-\beta_t\Bigl(\beta_0^2k+m+{mv^2\over 2}\Bigr)\Bigr]
\end{eqnarray}
Despite appearances, $\hat G_S(k)$ and $\hat L_S(k)$ are
finite at $k=m$ and $v=0$.

        Eq.(4.6) is algebraically complicated but is solvable by
conventional Wiener-Hopf methods.  We start by writing $\hat G_S(k)$
in the form $\hat G_S^{(+)}(k)\,\hat G_S^{(-)}(k)$, where the
superscripts $(\pm)$ have their usual significance.  We then
divide both sides of (4.6) by $\hat G_S^{(-)}(k)$ and rewrite the
result in the form:
\begin{eqnarray}
\nonumber
&& \left[{1\over \hat G_S^{(-)}(k)}\right]\,\hat{\bar\Sigma}_{S1}^{(-)}(k)
-{i\,\hat E_m\over k-m+i\epsilon}\,\left[{1\over \hat G_S^{(-)}(k)}\right]
+{i\,\hat E_m\over k-m-i\epsilon}\left[{1\over \hat G_S^{(-)}(k)}
-{1\over \hat G_S^{(-)}(m)}\right] \\
&&-\hat\Lambda_m^{(-)}(k)\hat Y_m
=-\hat G_S^{(+)}(k)\,\hat U_{S1}^{(+)}(k)
-{i\,\hat E_m\over k-m-i\epsilon}\,\left[{1\over \hat G_S^{(-)}(m)}\right]
+\hat\Lambda_m^{(+)}(k)\hat Y_m  .
\end{eqnarray}
The terms proportional to $\hat E_m$ come from writing $\delta (k-m)$ as the
difference between poles at $k=m\pm i\epsilon$, $\epsilon\to +0$, and then
writing the product $\bigl[(k-m-i\epsilon)\hat G_S^{(-)}(k)\bigr]^{-1}$
 as a sum of
terms each with singularities only in the upper or lower half $k$-plane.
Similarly,
\begin{equation}
\hat\Lambda_m^{(\pm)}(k)=\mp\int_{C^{(\pm)}}{dk'\over 2\pi i}\,
{1\over k'-k}\,\left[{\hat L_S(k')\,\hat U_{N0}^{(+)}(k'-m)\over
\hat G_S^{(-)}(k')}\right],
\end{equation}
where the contours $C^{(+)}$ and $C^{(-)}$ go from $-\infty$ to $+\infty$
in the $k'$ plane passing, respectively, above and below the pole at $k=k'$.
All of the terms on the left-hand side of (4.9) are $(-)$ terms and
those on the right are $(+)$; therefore both sides must be equal to
the same entire function of $k$.

        Far ahead of the crack tip, $x\to -\infty$, the shear stress on the
centerline $Y_{cen}(x)$ is $\hat E_m \exp(imx)$; thus
\begin{equation}
\displaystyle\lim_{k\to m}(k-m)\,\hat{\bar\Sigma}_{S1}^{(-)}(k) =i\hat E_m.
\end{equation}
It is convenient to eliminate the singularity in (4.9) by multiplying through
by $(k-m)$.  (In $x$ space, we are taking a derivative.) The resulting equation
has no simple poles near the real $k$ axis.  Moreover,
\begin{equation}
\displaystyle\lim_{k\to m}(k-m)\,\hat\Lambda_m^{(-)}(k)=0.
\end{equation}
To see this, note that the factor $\hat U_{N0}^{(+)}(k'-m)$ in the integrand in
(4.10) does have a simple pole at $k'=m+i\epsilon$ with a residue
proportional to the opening displacement $\varepsilon_{\infty}^{(2)} W$.
The contour $C^{(-)}$ passes below
both this pole and the one at $k'=k$, thus $\hat\Lambda_m^{(-)}(k)$ has no
singularity when $k\to m$.  On the other hand, the contour $C^{(+)}$ is pinched
between $k'=m+i\epsilon$ and $k'=k$ as $k\to m$, thus
$\hat\Lambda_m^{(+)}(k)$ does have a pole here.

        Eqs.(4.11) and (4.12) tell us that, after multiplication by $(k-m)$,
the entire left-hand side of (4.9) vanishes as $k\to m$, which
means that both sides of this
equation vanish for all $k$.  Accordingly, the Wiener-Hopf solutions are:
\begin{equation}
i(k-m)\,\hat{\bar\Sigma}_{S1}^{(-)}(k)
=-\hat G_S^{(-)}(k)\left[{\hat E_m\over \hat G_S^{(-)}(m)} -i(k-m)\,
\hat\Lambda_m^{(-)}(k)\hat Y_m   \right];
\end{equation}
\begin{equation}
i(k-m)\hat U_{S1}^{(+)}(k) =
{1\over \hat G_S^{(+)}(k)}\left[{\hat E_m\over \hat G_S^{(-)}(m)} +i(k-m)\,
\hat\Lambda_m^{(+)}(k)\hat Y_m   \right].
\end{equation}

\section{First-Order Calculations: Singularities in the Shear Stress}

        To interpret these results, it is useful to look first at the static
limit, $v\to 0$, and also to let $\hat Y_m   = 0$.  That is, we use (4.13) to
compute the shear stress
near the tip of a semi-infinite straight crack in a strip subject to the same
tractions --- including the small oscillating shear --- that we have been
considering so far.  Define
\begin{equation}
\bar\Sigma_{S1}(x,v=0,Y_{cen}=0)\equiv s_m(x)\,e^{imx}; 
\end{equation}
and note that
\begin{equation}
\displaystyle\lim_{v\to 0}\,\hat G_S(k)= \bigl(1-{1\over\kappa}\bigr)|k|
=\displaystyle\lim_{\epsilon\to 0}\bigl(1-{1\over\kappa}\bigr)\,
\bigl(\epsilon+ik\bigr)^{1/2}\bigl(\epsilon-ik\bigr)^{1/2}.
\end{equation}
The appropriate factorization of $\hat G_S(k)$ is obvious and, from (4.13), we
find for $x<0$:
\begin{eqnarray}
\nonumber
{ds_m(x)\over dx}
&&=-\displaystyle\lim_{\epsilon\to 0}{\hat\varepsilon_m}\int{dk\over 2\pi}
\left[\epsilon-
i(k+m)\over\epsilon-im\right]^{1/2} e^{ikx} \\
&&=-{|{\hat\varepsilon_m}|\,e^{-im(x+x_0)}\over2\sqrt{-i\pi m}|x|^{3/2}}
\end{eqnarray}
where we have defined the phase of $\hat \varepsilon_m$
by writing $\hat\varepsilon_m=|\hat\varepsilon_m|
\exp(- imx_0)$. Thus, the singular part of the static shear stress
near the tip is
\begin{equation}
\bar\Sigma_{S1}\approx {|{\hat\varepsilon_m}|\over\sqrt{\pi m |x|}}\,
\cos\bigl(mx_0-{\pi\over 4}\bigr).
\end{equation}
(A similar result has been obtained by Gao\cite{Gao}.)  We see here the usual
inverse-square-root stress singularity, but the half wavelength $\pi/m$
replaces the macroscopic length that ordinarily appears in the prefactor.

        The same kind of stress singularity occurs for finite $v$ if we
consider only a straight crack by setting $\hat Y_m  =0$ in (4.13).  The
full kernel $\hat G_S(k)$ diverges like $|k|$ for large $k$, and it is
this divergence that controls the stress
singularity near $x=0$.  As stated in the Introduction, our strategy is to
compute the amplitude $\hat Y_m  $ by choosing it in such a way as to
cancel this singularity.  Luckily, we do not need to carry out a
complete calculation of $\bar\Sigma_{S1}$ to find $\hat Y_m  $.

        In analogy to (5.1), we write
\begin{equation}
\bar\Sigma_{S1}(x)=S_m(x)\,e^{imx}
\end{equation}
and then separate $S_m$ into two parts, say $S_{m1}$ and $S_{m2}$,
corresponding to the two terms on the right-hand side of (4.13).  That is,
\begin{equation}
{dS_{m1}\over dx}=-{\hat E_m\over \hat G_S^{(-)}(m)}\,\int{dk\over 2\pi}\,\hat
G_S^{(-)}(k+m)\,e^{ikx}
\end{equation}
and
\begin{equation}
S_{m2}(x)=\hat Y_m  \,\int{dk\over 2\pi}\,\hat G_S^{(-)}(k+m)\,
\hat\Lambda_m^{(-)}(k+m)\,e^{ikx}.
\end{equation}
Using the definition of $\hat\Lambda_m^{(-)}$ in (4.10), we can
rewrite (5.7) as follows:
\begin{eqnarray}
\nonumber
S_{m2}(x)&&={\hat Y_m  \over\hat E_m}\int_{-\infty}^x dx'\,
{dS_{m1}(x')\over dx'}\,
\int_0^{\infty}dx''\,\Xi_m(x'+x''-x)\,U_{N0}(x'') \\
&&\approx {\hat Y_m  \over\hat E_m}\,S_{m1}(x)
\int_0^{\infty}dx'\,\Xi_m(x')\,U_{N0}(x')
+(\hbox {terms regular at}\,\,x=-0);
\end{eqnarray}
where
\begin{equation}
\Xi_m(x)= \hat G_S^{(-)}(m)\,\int{dk\over 2\pi}\left[{\hat L_S(k)\over
\hat G_S^{(-)}(k)}\right]e^{-i(k-m)x}.
\end{equation}

        Having written $S_{m2}(x)$ in this form, it is easy to see that
cancellation of the singularity in $S_{m}(x)$ requires:
\begin{equation}
\hat E_m = - \hat Y_m  \int_0^{\infty}dx\,\Xi_m(x)\,U_{N0}(x),
\end{equation}
or, equivalently,
\begin{equation}
\hat Y_m  =-{{\hat\varepsilon_m}\over im\Delta\epsilon_{\infty}+
\tilde{\cal D}(v,m)},
\end{equation}
where,
\begin{equation}
\tilde{\cal D}(v,m)= \int_0^{\infty}dx\int {dk\over 2\pi}\,
\left[\hat G_S^{(-)}(m)\,\hat L_S(k)\over
\hat G_S^{(-)}(k)\right]\, e^{-i(k-m)x}\,U_{N0}(x),
\end{equation}
and we have used the definition of $\hat E_m$ in (4.5).  We can
now use the representation of $U_{N0}(x)$ in (3.17) to find:
\begin{equation}
\tilde{\cal D}(v,m)=\varepsilon_{\infty}^{(2)}
\left[\kappa'W \over 2b(v)\,i^3\right]^{1/2}\int{dk \over
2\pi}\,{\hat G_S^{(-)}(m)\,\hat L_S(k)\over \hat G_S^{(-)}(k)\,
(k-m-i\epsilon)^{3/2}}.
\end{equation}
The infinitesimal quantity $\epsilon$  reminds us that the contour of
integration
passes below the singularity at $k=m$. We have made one approximation in
(5.13): The factor $(k-m-i\epsilon)^{3/2}$ in the denominator is, more
accurately, the product
\begin{equation}
(k-m-i\epsilon)\left(k-m-i\epsilon - {i\kappa'\over 2Wb(v)}\right)^{1/2};
\end{equation}
but we are required by previous approximations to take the limit
$W\to\infty$ in terms of this kind.  As a result, the only length scale
in (5.13) apart from the $\sqrt W$ in the prefactor, is the inverse
wavenumber $m^{-1}$. $\hat L_S(k)$ has dimension $m^2$, thus we can
extract a factor $m^{3/2}$ from the integrand in (5.13) and write:
\begin{equation}
\tilde{\cal D}(v,m)=im\varepsilon_{\infty}^{(2)} (-imW)^{1/2} \,
{\cal D}_Y(v). 
\end{equation}
We have now recovered Eq.(1.5) with
\begin{equation}
{\cal D}_Y(v)=-{1\over m^{3/2}}\left[\kappa'\over 2b(v)\right]^{1/2}
\int{dk\over
2\pi}\,{\hat G_S^{(-)}(m)\,\hat L_S(k)\over \hat G_S^{(-)}(k) \,
(k-m-i\epsilon)^{3/2}}, 
\end{equation}
which is an $m$-independent function of $v$.

\section{Evaluation of ${\cal D}_Y$}

        It remains to evaluate ${\cal D}_Y(v)$.  This is a tedious
process that requires some nontrivial analysis.  For completeness, we
outline the calculation here and record some essential formulas.
Readers not interested in the technical details should skip to the
conclusions in Section VII.

        The first step is to compute the factors $\hat G_S^{(\pm)}(k)$.
Let $k=mp$ and then write $\hat G_S(k)$ in the form:
\begin{equation}
\hat G_S(mp)={2m\over v^2}\,\left(p\over p-1\right)^2\,Q_l(p)\,
\left[1-{Q_0^4(p)\over p^2\,Q_l(p)\,Q_t(p)}\right]
\end{equation}
where
\begin{equation}
Q_0^2(p)\equiv {q_0^2\over m^2}= \beta_0^2\,(p-p_{0+})\,(p-p_{0-});
\qquad p_{0\pm}=\pm{v\over \sqrt 2 \pm v}\pm i\epsilon;
\end{equation}
\begin{equation}
Q_l^2(p)\equiv {q_l^2\over m^2}= \beta_l^2\,(p-p_{l+})\,(p-p_{l-});
\qquad p_{l\pm}=\pm{v\over \sqrt\kappa \pm v}\pm i\epsilon;
\end{equation}
\begin{equation}
Q_t^2(p)\equiv {q_t^2\over m^2}= \beta_t^2\,(p-p_{t+})\,(p-p_{t-});
\qquad p_{t\pm}=\pm{v\over 1 \pm v}\pm i\epsilon;
\end{equation}
Placement of the branch points relative to the real axis in the
complex $p$ plane
is determined by the conditions following Eq.(2.13).

        In the limit $p\to\infty$, the quantity in square brackets in
(6.1) goes to
\begin{equation}
B(v) = 1-{\beta_0^4\over \beta_l\beta_t}={v^2\over 2\beta_t}\,b(v).
\end{equation}
Accordingly, we remove a factor $B(v)$ from the square brackets and write:
\begin{equation}
\hat G_S(mp)={2m\over v^2}\,B(v)\,Q_l(p)\,
\exp\left[M^{(+)}(p)+M^{(-)}(p)\right]
\end{equation}
where
\begin{equation}
M^{(\pm)}(p)=\mp\int{dp'\over 2\pi i}\,{1\over (p'-p\pm i\epsilon)}
\ln\left\{{1\over
B(v)}\,\left(p'\over p'-1\right)^2\,\left[1-{Q_0^4(p')\over
p'^2\,Q_l(p')\,Q_t(p')}\right]\right\}.
\end{equation}
The logarithm in (6.7) vanishes at large $p'$ --- thus we can
close contours at infinity in the complex $p'$ plane --- and its
only singularities are at the branch
points of the functions $Q_0$, $Q_l$, and $Q_t$.
The factor $[p'/(p'-1)]^2$ in the
argument of the logarithm cancels singularities at $p'=0$ and $p'=1$.
Moreover, one can check that this argument has no zeroes on the
physical sheet. To evaluate (5.15), we need only $M^{(-)}(p)$, which
we can obtain from (6.7) by closing the contour around the branch cut
extending from $p_{t-}$ to $p_{l-}$ just below the negative real axis.
The result is:
\begin{equation}
M^{(-)}(p)={1\over\pi}\int_{p_{t-}}^{p_{l-}}{dp'\over p-p'}\,
\theta(p') 
\end{equation}
where
\begin{equation}
\tan\theta(p)={\beta_0^4(p-p_{0+})^2(p-p_{0-})^2\over
\beta_l\beta_tp^2\left[(p-p_{t-})\,(p_{t+}-p)\,(p_{l-}-p)\,
(p_{l+}-p) \right]^{1/2}}
\end{equation}
Then, the needed factor of the Wiener-Hopf kernel is:
\begin{equation}
\hat G_S^{(-)}(pm)=(p-p_{l-})^{1/2}\,\exp\left[M^{(-)}(p)\right].
\end{equation}

        Carrying out the remaining integration in (5.15) is
primarily a chore of figuring out how to close contours
around the various branch cuts in the complex $p$-plane.
Our results are the following:
\begin{equation}
{\cal D}_Y(v)={2\over \pi v^2}\left[\kappa'\,(1-p_{l-})\over
2b(v)\right]^{1/2}\,e^{R(1)}\,\sum_{n=1}^3\,D_n(v),
\end{equation}
where the three terms in the sum correspond to the three terms
in $\hat L_S$ shown in (4.8), and
\begin{equation}
R(p)=\wp\,\int_{p_{t-}}^{p_{l-}}{dp'\over \pi}\,{\theta(p')\over p-p'}.
\end{equation}
The symbol $\wp$ denotes a Cauchy principal-value integral. We find:
\begin{mathletters}
\begin{equation}
D_1(v)=\beta_l\,\int_{p_{t-}}^{p_{l-}}dp\,\,
{p(p_{l+}-p)^{1/2}\over(1-p)^{3/2}}\, e^{-R(p)}\,
\sin\theta(p)\,f_1(p);\label{equationa}
\end{equation}
\begin{equation}
f_1(p)= \beta_0^2-\left(1\over 1-p\right)^2;\label{equationb}
\end{equation}
\end{mathletters}
\begin{mathletters}
\begin{equation}
D_2(v)= {\beta_0^2\over\beta_t}\,\int_{p_{t-}}^{p_{l-}}dp\,\,
{(p-p_{0+})(p-p_{0-})\,e^{-R(p)}\,\cos\theta(p)\,
f_2(p)\over\left[(p-p_{t-})(p_{t+}-p)
(p_{l-}-p)\right]^{1/2}(1-p)^{3/2}};\label{equationa}
\end{equation}
\begin{equation}
f_2(p)={p\over(1-p)^2}-p-v^2(1-p);\label{equationb}
\end{equation}
\end{mathletters}
\begin{mathletters}
\begin{equation}
D_3(v)=\int_{p_{t-}}^{p_{l-}}dp\,\,{e^{-R(p)}\,
\sin\theta(p)\,f_3(p)\over(1-p)^{1/2}
(p_{l-}-p)^{1/2}}+\int_{p_{l-}}^1 dp\,\,
{e^{-R(p)}\,f_3(p)\over (1-p)^{1/2}(p-p_{l-})^{1/2}};\label{equationa}
\end{equation}
\begin{equation}
f_3(p)= \beta_t\left(\beta_0^2p+1+{v^2\over2}\right) -
{\beta_0^2\over\beta_l}\left(\beta_l^2p+1\right).\label{equationb}
\end{equation}
\end{mathletters}
These quantities are now in a form suitable for numerical evaluation.

\section{Concluding Remarks}

        In Fig.1, we show ${\cal D}_Y$ as a function of $v/v_R$
for several different values of $\kappa$.  We have
chosen $\kappa'=\kappa$, that is, $\varepsilon_{\infty}^{(1)}=0$.
As advertised in Section I,
${\cal D}_Y(v)$ is positive at small
$v$ and decreases through zero at $v=v_c(\kappa)$ with $v_c$
somewhat less than the Rayleigh speed $v_R$.  The divergence
at $v=v_R$ is artificial; we have assumed at various places
in this analysis that the product $Wb(v)$ is
large, but this assumption breaks down at $v_R$ where $b(v)$ vanishes.

Clearly, the system undergoes some kind of loss of stability at $v_c$.  The
amplitude of the oscillation of the crack jumps from a small value of order
$W^{-1/2}$ for $v<v_c$ to $i\,{\hat\varepsilon_m}/(m\,\Delta\epsilon_{\infty})$
at $v=v_c$. At precisely $v_c$, the crack is following the curve along which
the shear stress vanishes in the unbroken material.  An alternative way to
interpret  this situation is to note that the quantity
$\varepsilon_{\infty}^{(2)}\sqrt W$ is  proportional to the stress-intensity
factor and therefore is a $W$-independent  function of the velocity $v$.  From
this point of view, the response to the applied  shear stress for $v<v_c$ is
independent of the macroscopic size of the system but,  because
$\Delta\epsilon_{\infty}$ is intrinsically a very small quantity, the response
becomes very large at $v=v_c$.

        No details of the actual dynamical model such as the
cohesive stress or the viscous dissipation at the fracture
surfaces seem to be relevant to these results.  In
calculating ${\cal D}_Y$, we use the fact that the zero'th order
velocity $v$ is a well defined function of the applied stresses,
but neither $\Sigma_c$ or $\eta$ enter the final formulas in
any explicit way.  This situation would change if we were
interested in values of the perturbing wavelength
$2\pi/m$ so small that they were comparable to the size of the
cohesive zone, in which case (3.17) would not be an adequate
approximation for the crack opening displacement $U_{N0}(x)$.
The question of whether interesting
instabilities might occur on such short length scales is one
of the issues that we must leave for later investigation.

        We see no mathematical relationship between our results
and the usual analysis of the Yoffe instability. Our formulas
do not reproduce the Yoffe criteria for a shift in the maximum
of the circumferential stress or a shift in the direction
along which the shear stress vanishes.  We suspect that the latter
calculation is closest to ours, the difference being that we
are computing the shear stress near the tip of a moving wavy
crack instead of a straight one.  Our results indicate that,
by shifting to a very slightly (order $W^{-1/2}$) wavy trajectory
in response to an oscillatory perturbation, the crack regularizes
the shear stresses near its tip up to speeds appreciably higher
than those predicted previously.

        The fact that $\hat\chi_Y$ in (1.5) remains finite
even at $v_c$ means that
this system does not undergo a true linear instability, at
least not in response to
a shearing perturbation.  (The only divergence in
$\hat\chi_Y$ occurs at $v_c$ when $\Delta\epsilon_{\infty}=0$,
that is, when the stresses in the unperturbed system are
rotationally symmetric.)  A closer look reveals that the system
is just neutrally
stable.  In fact, neutral stability follows from very
general considerations.  In the limit of large $W$, the
location of the $x$-axis must be irrelevant; thus, if an
otherwise stable running crack is displaced by some perturbation
from one line parallel to the $x$-axis to another, it must
simply stay there, neither returning to
its original position or moving further away from it.

        We can see this explicitly.  Because we have
computed $\hat Y_m  $ just to
linear order in ${\hat\varepsilon_m}$, we can superimpose
these solutions to compute the
response to a sharp pulse of shear stress.  That is, we
let ${\hat\varepsilon_m}$ equal some
$m$-independent quantity, say $\gamma$, so that
\begin{equation}
\varepsilon_{shear}(x)=\int{dm\over 2\pi}\,
{\hat\varepsilon_m} \,e^{imx}=\gamma\,\delta(x).
\end{equation}
Strictly speaking, we must make the width of this pulse much
greater than the length of the cohesive zone and much less
than $W$ in order to be consistent with our approximations.
Also, we cannot take (1.5) literally for values of $m$
such that $mW$ is of order unity or smaller.  With these
reservations, we write:
\begin{equation}
{dY_{cen}\over dx}=\gamma\,\int{im\,dm\over2\pi}\,
\hat\chi_Y(m,v)\,e^{imx}.
\end{equation}
For $v<v_c$, the integrand in (7.2) is dominated by the
term containing $(mW)^{1/2}$ in the denominator of
$\hat\chi_Y$, and thus we find that the initial
response to this perturbation is
\begin{equation}
Y_{cen}(x)\approx {2\gamma\over\varepsilon_{\infty}^{(2)}
{\cal D}_Y(v)} \,
\left(-\pi x\over W\right)^{1/2}
\end{equation}
for $x$ small and negative.  Note that this initial response is
suppressed by the
factor $W^{-1/2}$.  Unfortunately, we cannot determine the ultimate
displacement of the crack from (7.2) because, to do so, we
would need to know
$\hat\chi_Y$ for small values of $mW$.  The perturbed crack
tip apparently moves a distance of order $W$ along the $x$-axis
before coming to its new
steady-state position and, therefore, must feel the boundaries
of the system in a
way that is not included in our analysis.

        At $v=v_c$, the situation is markedly different.
The integrand in (7.2) is
independent of $m$, thus the integration produces a $\delta$
function that means that the crack jumps abruptly to its final
displacement, $\gamma/\Delta\epsilon_{\infty}$.
The crack tip ceases to be controlled by the stress field
generated by the semi-infinite fracture that it has left behind and,
instead, responds instantaneously to
perturbations in its path.  Interestingly, the direction in which it
jumps is determined by the sign of $\Delta\epsilon_{\infty}$.

At this point in the discussion, it is logical to ask whether our results might
be relevant to the laboratory and numerical  experiments mentioned at the
beginning of Section I.  Clearly, we are far from solving the problem.  Our
critical velocity $v_c$ is appreciably higher than the onset speeds reported by
Fineberg {\it et al.}\cite{Texas}.  Moreover, those experiments suggest that
there is some characteristic wavelength associated with the onset of
instability, whereas our formula (1.5) associates no special value of $m$ with
the change in behavior at $v_c$.  On the other hand, this change in behavior at
$v_c$ appears to be a general dynamic property of mode-I fracture, and it seems
unlikely  to be completely irrelevant.

There are at least three possibilities.  One is that we simply have chosen the
wrong kind of perturbation. We are presently looking to see  whether some other
oscillatory term in (2.4) might destabilize the crack at slower speeds.  A
second possibility is that our wavy crack itself becomes unstable against some
new perturbation at $v<v_c$.  We can imagine, for  example, a coupling between
wavy modes and perturbations that break the two- dimensional symmetry.  A third
important possibility is that nonlinear effects are  playing a role. Our
critical velocity $v_c$ appears to be some kind of bifurcation  point. The
solutions for $v>v_c$ must be unstable.  We have not shown this  explicitly,
but the fact that the crack tip moves in a direction opposite to that of the
perturbing shear when ${\cal D}_Y(v) <0$ in (7.3) may indicate that  these
states are not physically realizable. If this were a more familiar kind of
dynamical system, we would conclude that the change in stability at $v_c$
implies the emergence of one or more branches of stable states at that point.
Those states would be accessible mathematically only {\it via} nonlinear
analysis.  Pushing the analogy further, we might guess that we are dealing with
a subcritical bifurcation at $v_c$, so that the system might be driven into a
strongly oscillatory mode by nonlinear perturbations at values of $v$ less than
$v_c$.  If this  is the correct picture, then it might be possible to achieve
higher critical crack speeds by using very clean, defect-free samples in the
experiments.

\acknowledgements

We thank M. Leibig for doing the computations required for
preparing Figure 1.  This research was supported by U.S. DOE
Grant No. DE-FG03-84ER45108 and NSF Grant No. PHY89-04035.

\bigskip

\centerline{FIGURE CAPTION}
Fig. 1.  ${\cal D}_Y$ as a function of $v/v_R$ for $\kappa = $2 (solid line), 3
(short dashed line), and 4 (long dashed line).

\end{document}